# Millimeter-scale rigid diamond probe for high sensitivity endoscopic-magnetometry applications

Jihongbo Shen[1], Heng Yuan[1,*], Hongyu Tao[1], Zekun Niu[1], Haoming Xu[1], Chentao Zhang[1], Chen Su[1], Zhuo Wang[1], Chen Zhang[2,*]

**AFFILIATIONS**

[1]*School of Instrumentation and Optoelectronic Engineering, Beihang University, Beijing 100191, China*

[2]*Beijing Quantum Technology Research and Development Center, Cuiyunxi Rd. 1, Beijing 100074, China*

[*]Corresponding authors:
hengyuan@buaa.edu.cn (Heng Yuan)
chen.zhang@cquterd.cn (Chen Zhang)

## Abstract

Magnetometry based on diamond nitrogen-vacancy (NV) centers has been extensively studied for applications requiring diverse capabilities, spanning from nanometer spatial resolution to subpicotesla sensitivity. Among various applications, diamond magnetometers can demonstrate high sensitivity magnetic sensing within millimeter-scale size for endoscopic applications. However, the trade-off between sensitivity and spatial resolution of diamond magnetometry makes it difficult to achieve such a probe. In this study, we present a millimeter-scale rigid diamond magnetometer probe with enhanced sensitivity via optimizing the optical design. By coupling the frustum diamond with the miniaturized compound parabolic concentrator (CPC) lens, we enhance the fluorescence collection efficiency by 37% within 4 mm diameter, and the achieved sensitivity is 200 pT/Hz$^{1/2}$ based on the sample with the resonance linewidth of ~8 MHz. With this verified structure, endoscopes with mm-size probe and picotesla sensitivity can be projected for surgical and industrial applications in the future.

## 1. Introduction

The negatively charged nitrogen-vacancy centers (NV) in diamond exhibit great potentials一e.g. high sensitivity [1], [2], [3], high spatial resolution [4], [5] and large dynamic range [6], [7]一for various magnetometry applications. It has emerged as promising quantum sensors for analyzing materials [4], [5], monitoring electricity networks [8], [9], detecting biomagnetic fields [10], [11], etc. Conventionally, diamond magnetometry is developed with either extremely high spatial resolution or high sensitivity for different applications. For example, nano-diamonds are usually used for sub-cellular sensing [12], [13], and the scanning probe system is applied for analyzing materials with nanoscale resolution [14]. High sensitivity diamond magnetometers are developed with package size of centimeters for sensing tiny currents [15], etc. Despite that mm-size diamonds are usually used in high-sensitivity sensing scheme, the spatial resolutions of such sensors are still limited due to the optics for collecting fluorescence photons. Fiber probes are developed to improve the spatial resolution for magnetometers based on bulk diamonds [16], [17], [18], [19]. However, the sensitivity deteriorates a lot due to the poor fluorescence collection efficiency of fibers. Therefore, it is crucial to enhance fluorescence collection efficiency for diamond probes to keep high sensitivity in the package of a few millimeters. Compared with the flexible probes based on fibers, rigid probes based on optical guides made of glass can keep high sensitivity while miniaturized due to high fluorescence collection efficiency. Such diamond probes applied for endoscopic-magnetometry applications could detect weak magnetic signals in limited space in the future, e.g., magnetoneuro-/ magnetomyro-signals in digestive tract [13] (for example, in throat as shown in Fig. 1(a)) and magnetoencephalo-signals from skull base [12], which might encourage new science and surgical techniques.

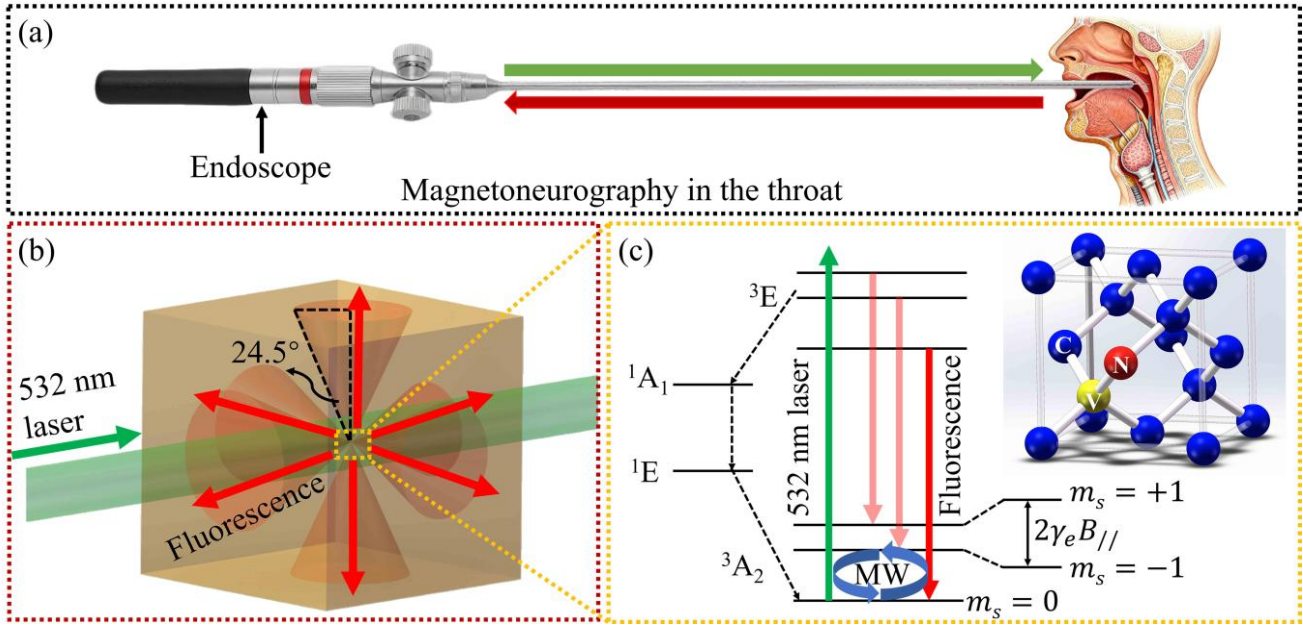

**Fig. 1.** (a) The application of the diamond probe, which shows the potential application of endoscopes in magnetoneurography. (b) The red box shows the schematic diagram of fluorescence propagation in the cube diamond. The fluorescence is emitted outside within a conical solid angle of only 24.5°. (c) The lattice structure of NV centers and the energy level structure are shown in the orange box. The ground states ($^3A_2$) are used for magnetic field measurement. Magnetic fields can be detected using microwave frequency sweeping.

Other cutting-edge magnetometers with high sensitivity also meet the challenge of reducing package size for endoscopic applications. For example, the dimension of advance optically pumped magnetometers [20], [21] can be above centimeter scale while keeping the sensitivity ranging from pT to fT, but the dimension could hardly decrease further due to the size of the vapor cell and other components. On the other hand, a sub-mm$^3$ high quality bulk diamond is already sufficient for achieving a sensitivity of a few pT [15], and the essential components of the endoscopic probe include an optical guide and a microwave antenna. The simple schematic makes it possible to build an endoscope with a few mm diameter.

The miniaturization of the endoscopic probe mainly depends on the design of the fluorescence (FL) collection optics, as magnetometer sensitivity scales directly with collected FL photons. FL collection optics, such as total internal reflection (TIR) lenses [22], three-way fluorescence collection structures (EC) [23], fluorescence omni-directional reflection concentrators (FORC) [24], concave mirrors [25], and commercial compound parabolic concentrator (CPC) lenses [26] have been explored to enhance FL collection efficiency. However, these optics exhibit centimeter-scale or larger diameters and are not used for probe miniaturization. Therefore, optimizing optics is essential for miniaturized and high sensitivity endoscopic-magnetometers.

In this study, we present a millimeter-scale rigid diamond probe by optimizing the optical design to enhance the sensitivity and minimize the size. This probe is applicable for high sensitivity endoscopic-magnetometry. By combining the frustum diamond and the optimized CPC lens, we demonstrate that the FL collection intensity increases by 37% within a 4-mm diameter dimension. The designed diamond probe achieves a 200 pT/Hz$^{1/2}$ sensitivity with an optically detected magnetic resonance (ODMR) linewidth of ~8 MHz. This simple and validated configuration enables the development of millimeter-scale endoscopic probes capable of picotesla-level sensitivity, which is expected to provide strong support for various applications like surgical interventions and industrial diagnostics in the future [27].

## 2. Methods

*2.1 Analysis of fluorescence emission*

The fluorescence emission pathways of the cube diamond sample upon excitation with a 532 nm laser are illustrated in Fig. 1(b). In the continuous-wave (CW) ODMR scheme, the sensitivity ($\eta$) of the diamond magnetometer is related to the number of collected fluorescence photons ($R_{fl}$) involved in detection by a factor of fluorescence $\sqrt{R_{fl}}$, as is described below [15]:

$$\eta \approx P_F \frac{h}{g\mu_B} \frac{\Delta\nu}{C\sqrt{R_{fl}}} \tag{1}$$

where $\Delta\nu$ is the linewidth of the ODMR spectrum, $C$ is the signal contrast, $P_F$ is a factor related to the line shape, $h$ is the Planck constant, $g$ is the electron g-factor and $\mu_B$ is the Bohr magneton. Thus, enhancing total collected fluorescence ($TCF$, $TCF \propto R_{fl}$, detailed analysis is in Supplementary Material 1) by optical designs improves the magnetic sensitivity of the diamond magnetometer. Using the fluorescence generated inside a cube diamond as the reference, the total collected fluorescence ($TCF$) is defined as the ratio of the total fluorescence collected by the PD to the fluorescence generated inside the cube diamond. The total emitted fluorescence ($TEF$) is defined as the ratio of the fluorescence emitted from the diamond to the fluorescence generated inside the cube diamond, where a portion of the emitted fluorescence is ultimately not collected by the PD. Additionally, the fluorescence emitted efficiency ($\eta_e$) is defined as the ratio of the fluorescence emitted from the diamond to the fluorescence generated inside the diamond of its current shape (i.e., the frustum diamond). During the processing of a cube diamond into a frustum shape through cutting, some volume is lost, leading to a reduction in the number of NV centers and consequently a decrease in the internally generated fluorescence. To rigorously evaluate the $TCF$ improvement resulting from the change in diamond shape, the volume must be taken into account when analyzing the performance of the processed frustum diamond ($TEF = V_s/V_c \times \eta_e$, $TCF = TEF \times \eta_c$; $V_s$, $V_c$, and $\eta_c$ represent the volume of frustum diamond, the volume of cube diamond, and the collection efficiency of optical lenses, respectively; a detailed analysis is provided in Supplementary Material 1).

Due to the high refractive index of diamond ($n = 2.41$), the fluorescence generated by each NV center within the cube diamond only emits from the diamond within the six directions of the conical solid angle shown in Fig. 1. The semi-apex angle of this cone is 24.5°, corresponding to the critical angle for total internal reflection at the diamond-air interface. Based on this, the fluorescence emitted efficiency ($\eta_e = 27\%$) of the cube diamond is calculated using the solid angle model:
$$\eta_e = 6\Omega/\Omega_{\text{all}} = 3(1 - \cos\alpha) \tag{2}$$
where $\Omega = 2\pi(1 - \cos\alpha)$ refers to the solid angle decided by the conical semi-apex angle. $\Omega_{\text{all}} = 4\pi$ reflects the total sum of solid angles in all directions from a point within the cube diamond. Fluorescence emitted in other directions undergoes total internal reflection within the diamond and cannot emit out of the diamond. When the diamond is surrounded by optical adhesive (OA, $n = 1.51$) and its side surfaces are modified into inclined surfaces, the emitted fluorescence from the diamond increases. The analysis in detail is shown in Fig. S1 in the Supplementary Material.

Fig. 1(c) shows the crystal structure and the energy level diagram of the diamond NV centers. An NV center in diamond constitutes a point defect formed by a nitrogen atom substituting for a carbon atom adjacent to a vacancy in the diamond lattice [28]. This defect exhibits a unique electronic structure with distinct energy levels. Upon laser excitation, the NV center emits red fluorescence [29]. The intensity of this red FL corresponds to the population of the NV center in its electronic ground state $|m_s = 0>$ [30]. The ground state features a zero-field splitting (ZFS), resulting in energy levels for $|m_s = \pm 1>$ and $|m_s = 0>$ with a splitting value [31], which is one of the characteristic features used to identify the NV center and controlled by the microwave (MW). The electronic spin states of the NV center can be optically initialized and read out, with the fluorescence intensity differing depending on the spin state. Specifically, the fluorescence intensity is stronger when the center is in the $|m_s = 0>$ state compared to the $|m_s = \pm 1>$ states. In the presence of an external magnetic field ***B***, the Zeeman effect lifts the splitting of the $|m_s = \pm 1>$ states. Consequently, the transition frequencies from the $|m_s = 0>$ state to the $|m_s = -1>$ and $|m_s = +1>$ states become $f_-$ and $f_+$, respectively. If the magnetic field vector ***B*** forms an angle $\theta$ with the NV symmetry axis, the magnetic field projection along this axis is $B_{//} = B\cos(\theta)$. Thus, the external magnetic field magnitude along the NV axis is directly detected via shifts in the resonance frequencies observed in the optically detected magnetic resonance (ODMR) spectrum. The ODMR spectrum is acquired by sweeping the applied MW frequency and monitoring the fluorescence change [32].

*2.2 Design of diamond probe*

Enhancing the sensitivity of diamond magnetometer probes necessitates an increase in the number of collected photons [22]. This requires improvements in both the emission efficiency of fluorescence from the diamond and the collection efficiency of the emitted fluorescence. For the probe design, minimizing the probe's diameter while achieving a narrow divergence angle is essential to facilitate long-distance fluorescence transmission. A CPC lens is an effective collector that minimizes the divergence angle, thereby reducing fluorescence loss at the lens-air interface during transmission. Fig. 2(a) illustrates the optimization process for both the fluorescence (FL) collection lens and the diamond shape. The commercial CPC lens (material: B270, $n = 1.52$) exhibits a larger divergence angle and diameter, making it unsuitable for a slim probe. To address this, a short CPC lens (material: B270) is developed by reducing the length of the standard CPC lens (fitted with formula Eq. S4). This short CPC has half the length of the optimized CPC and one-third the length of the commercial CPC. The short CPC lens is also the commercial product. Further optimization yields an optimized CPC lens (material: high borosilicate glass, $n = 1.47$), achieving significant reductions in diameter (3 mm) and divergence angle (30° in air). However, due to the complex manufacturing process of CPC lenses, particularly when using high refractive index materials (e.g., H-ZLAF75B, $n = 1.90$), a tapered rod lens is explored as an alternative. The optimized tapered rod lens has a diameter of 3.4 mm and a divergence angle of 31.5°. While its fluorescence collection efficiency is lower than that of a low-refractive-index CPC lens, it offers structural simplicity and greater manufacturability. The diamond sample is a commercial and typical Type IIa (100)-oriented single-crystal diamond, obtained from bulk diamonds (Xianduan Crystal) through cutting and polishing processes, and is grown via chemical vapor deposition. The sample has a $^{14}$N concentration of <13 ppm and an NV concentration of 4.6 ppm after processing, and can be formed into a cube diamond with dimensions of 0.5 × 0.5 × 0.5 mm³. To enhance the diamond's emitted fluorescence efficiency, this study employs polishing and shaping of the cube diamond, as shown in the red box. However, this volume-reducing process leads to loss of NV centers and decreases total fluorescence generation. Therefore, it is crucial to consider the diamond volume. Additional optimization analysis is provided in Fig. S1 (Supplementary Material). The final optimized shape (FD-4 with a base angle of 86°) successfully enhances the diamond's emitted fluorescence efficiency while preserving more NV centers. After optimization, the emitted fluorescence efficiency reaches 81%, representing a three-fold improvement over the original cube diamond. The fluorescence enhancement effect of diamond sample geometry is analyzed purely from a geometric perspective, and the geometric design of FD-4 is suitable for most samples with uniform NV concentration distribution. Thus, it can be implemented in different types of samples. This study focuses on the influence of the base angle of the frustum diamond on $TCF$. Therefore, the angular error of the base angle is a key consideration in the manufacturing process, with a tolerance of <1° ($TCF$ loss is less than 1.4% compared to optimized base angle). All FD-4 structured samples used in the experiments were cut from the same larger diamond, ensuring the reproducibility of the results.

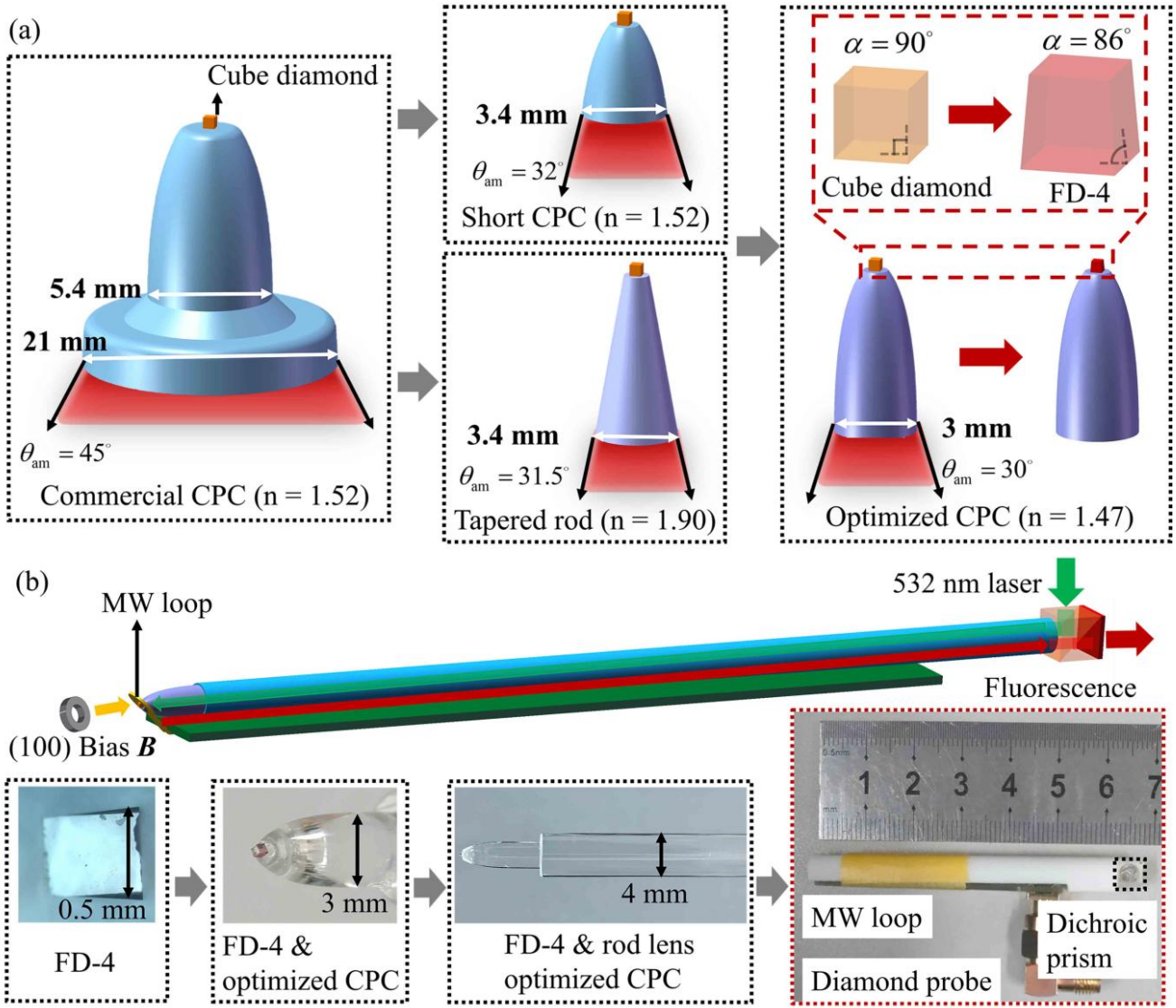

**Fig. 2.** (a) Optimization process of diamond shapes and FL collection lenses. The diamond is optimized from the cube to the frustum with 4 inclined planes (FD-4). The base angle ($\alpha$) is 86°. The commercial CPC lens is optimized for small output diameter ($d_{out}$) and divergence angle ($\theta_{am}$). The short CPC lens is derived from the standard CPC lens by reducing its length. The tapered rod lens is a substitute for the CPC lens due to its simpler fabrication process. (b) Schematic diagram and photograph of the diamond probe.

The diamond magnetometer structure combines a frustum diamond with 4 inclined planes, an optimized CPC lens, a rod lens (material: H-K9L, $n = 1.51$), a dichroic prism, a filter (Newsport, 10CGA-665), and a MW loop, as shown in Fig. 2(b). The design of the microwave loop is matched to the optical design size of the probe (4 mm diameter), with an impedance matching of 50 Ω. The applied magnetic field exhibits a uniformity of >95% within the central $0.5 \times 0.5$ mm² region (where the diamond is placed). For further details, please refer to Supplementary Material 8. The maximum MW power used from the MW source is 0 dBm. After passing through the power amplifier and circulator, the MW signal is applied to the MW loop. The microwave power delivered to the loop is approximately 30 dBm, which can cause heating of the loop. The frustum diamond is described in more detail in Fig. S2(a) in the Supplementary Material. The short cutting surface side length is 0.43 mm, decided by the base angle ($\alpha = 86°$). The base surface side length is 0.5 mm. It exhibits angled facets on the sides that effectively decrease the divergence angle of the diamond internal FL. This reduction in divergence angle greatly influences the efficient emission of a great amount of excited FL from the diamond and its subsequent capture by the lens. The optimized CPC lens is described in more detail in Fig. S2(b) in the Supplementary Material. The output diameter is 3 mm and the divergence angle is 30°. It can reduce the divergence angle of the emitted fluorescence and minimize FL loss from the lens side to the air. The smaller divergence angle (30°) contributes to the final focus on the small photosensitive area of the photodetector (PD). The rod lens achieves a long-distance FL transmission similar to optical fibers, with a larger diameter than the CPC lens (4 mm) in case of FL loss between the lenses. The larger diameter of the rod lens also makes it convenient for bonding. The dichroic prism is used to reflect the 532 nm laser and pass the FL from the rod lens. The laser incident surface and the fluorescence emission surface are perpendicular to each other. A long-pass filter is also added to obstruct the propagation of the laser, thereby minimizing laser interference with FL collection. The MW loop is used to conduct microwaves for the manipulation of the quantum state of NV centers. The circular ring design enables the diamond to be exposed to a more uniform microwave field. The ring-shaped permanent magnet is positioned along the [100] crystallographic direction of the probe to apply a bias magnetic field of 13 G (0.0013 T). All subsequent experimental results—including tests of fluorescence collection measurements, ODMR measurements, and sensitivity evaluations—are obtained under this [100]-oriented magnetic field environment. Static field along [100] axis projects equally onto all four NV classes, allowing all NVs to contribute to magnetometry signal.

## 3. Results and discussions

*3.1 Simulation results and optimized parameters*

The maximum $TCF$ comprehensively accounts for all design factors. The optimized parameters of all designs are shown in Tabel S1. Since $d_{out}$ reflects the overall maximum size of the magnetometer probe and $\theta_{am}$ determines the spot size of the FL onto the PD, the two parameters are the optimization parameters for the CPC lens [33]. The output radius $r_{out}$ is the half of the output diameter ($d_{out}$). The two parameters, $\theta_{am}$ and $r_{out}$, determine the surface shape of the CPC lens. The optimal parameters for the CPC lens are determined through simulation, with the diamond base surface length fixed at 0.5 mm. The root-mean-square radius ($RMSR$) is a statistical metric that describes the "width" or extent of energy spread of a light spot in space. It quantifies the dispersion of the spot's intensity distribution relative to its centroid (center of mass), which in this study corresponds to the center of the detector. In the simulation, the square of the root mean square radius ($RMSR^2$) is used to reflect the spot area, where the main optical intensity is concentrated on the detector [34]. The relationship among the $RMSR$, $r_{out}$, the distance between the probe and the PD ($d$), and $\theta_{am}$ is established in Fig. S8 of Supplementary Material 10 and is fitted from ray-tracing simulation results. The $RMSR$ reflects the $r_{out}$ and the $\theta_{am}$ according to:

$$RMSR \propto r_{out} + d\tan\theta_{am} \qquad (3)$$

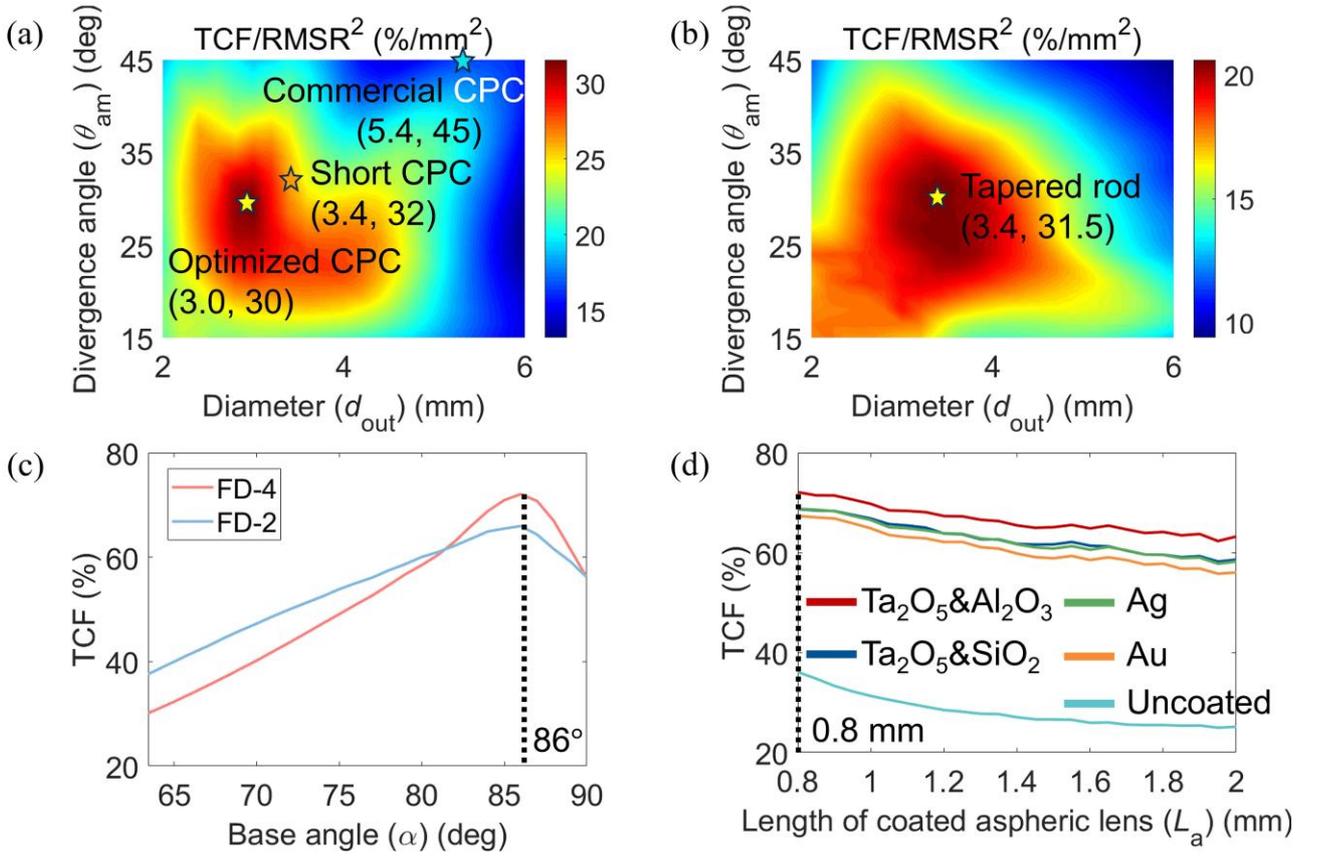

**Fig. 3.** Simulation results and optimization of diamond FL collection structures. (a) $TCF/RMSR^2$ for different CPC lens structures ($n = 1.47$). Root mean square radius ($RMSR$) reflects the radius of the spot radius on the PD. Larger size of the CPC lens makes more $TCF$ and more $RMSR$. (b) $TCF/RMSR^2$ for different tapered rod lens structures ($n = 1.90$). (c) The relationship between $TCF$ and $\alpha$ of the frustum diamond with 2 inclined planes (FD-2) and 4 inclined planes (FD-4). (d) The relationship between $TCF$ and the length of the coated aspheric lens under different coatings. The length of the aspheric lens ($L_a$) is above 0.8 mm for complete coverage.

where $d$ is the distance between the probe and the PD. Increasing $d_{\text{out}}$ and $\theta_{\text{am}}$ increases $TCF$ but increases the size of the CPC lens. Due to the increase in spot size with the increasing CPC dimension and the larger divergence angle, it is not conducive to the miniaturization of the CPC structure and the fluorescence conduction. Therefore, it is necessary to select the CPC total collected fluorescence per unit area ($TCF/RMSR^2$) as the CPC lens optimization target, as shown in Fig. 3(a). The commercial CPC lens has the smaller $TCF$ than the optimized CPC but larger $d_{\text{out}}$ and $\theta_{\text{am}}$, resulting in larger $RMSR$. The short CPC lens is designed only for miniaturization without optimizing $TCF$ and the diameter. The reduction in length of the short CPC lens increases the diameter and the divergence angle compared to the optimized CPC lens but has no effect on the $TCF$.

The optimization result of the tapered rod is shown in Fig. 3(b). The schematic diagram of the tapered rod is shown in Fig. S2(c). The tapered rod structure is similar to the CPC lens. Due to its simpler structure, it is easier to be fabricated. The tapered rod lens exhibits inferior collimation and FL collection efficiency compared to the CPC lens. The tapered rod lens is employed as an alternative to the CPC lens. For instance, in this study, High refractive index ($n = 1.90$) CPC lenses are hard to be fabricated and the tapered rod lens is utilized as a substitute. The purpose is to test the FL collection capability of collimating lenses with different refractive index. After optimization, the tapered rod shows the lager diameter and the divergence angle but lower $TCF$, as shown in Fig. 3(c).

The relationship between $TCF$ and diameters of these 4 optical designs is shown in Table 1. The optimized CPC lens shows the higher $TCF$ and smaller diameter and divergence angle.

**Table 1** Comparison of 4 optical designs.

| Optical designs | $TCF$ (%) | Diameter (mm) |
|---|---|---|
| Optimized CPC | 72.1 | 3.0 |
| Short CPC | 72.1 | 3.4 |
| Tapered rod | 67.8 | 3.4 |
| Commercial CPC | 70.6 | 5.4 |

$TCF$ is determined by varying $\alpha$ of the frustum diamond, as shown in Fig. 3(c). The $TCF$ is proportional to $V_s \times \eta_e$, where $V_s$ is the volume of the frustum diamond and $\eta_e$ is the fluorescence emitted efficiency of the frustum diamond. An increase in the base angle $\alpha$ decreases $V_s$ and increases $\eta_e$. The detailed analysis refers to the Supplementary Material 1. The maximum $TCF$ is achieved at 86°. The frustum diamond with 2 inclined planes (FD-2) has a larger volume than the frustum diamond with 4 inclined planes (FD-4) but has lower emitted efficiency. The $TCF$ of FD-4 demonstrates enhancement factors of 29% and 240% against the cube diamond and pyramid structures (PYD), respectively.

The coated aspheric lens is described in detail in Fig. S2(d). It can cover the diamond to collect the fluorescence from the other five faces of the diamond. The structure is determined by the length ($L_a$) and maximum radius ($R_a$). To ensure that the diamond is fully embedded, $R_a$ is sufficiently large but smaller than the input radius ($r_{\text{in}}$) of the CPC lens. The surface shape is optimized to be a parabolic surface. Fig. 3(d) shows the relationship between $TCF$ and $L_a$ under different coatings ($R_a = 0.5$ mm). Considering that the diamond is fully embedded, the minimum $L_a$ is 0.8 mm. A longer length results in smaller $TCF$ regardless of the coating. The reflectivity of the coating decides the $TCF$. The combination of tantalum pentoxide (Ta$_2$O$_5$) and aluminum oxide (Al$_2$O$_3$) shows the highest $TCF$ among the coatings. The combination of tantalum pentoxide (Ta$_2$O$_5$) and silicon dioxide (SiO$_2$) shows almost the same $TCF$ as the silver coating. The metal coating can absorb the FL and interfere with the magnetic field and MW field. Therefore, the non-metallic medium coating is a better choice to reflect the FL emitted from the diamond. The coated aspheric lens is not fabricated due to its complicated fabrication process. However, it offers the potential to further enhance FL collection and magnetometer sensitivity.

Detailed data on alignment tolerance analysis are provided in Supplementary Material 9. When bonding the three optical components—diamond, CPC lens, and rod lens—if the diamond base remains fully in contact with the input surface of the CPC lens without air gaps (translational misalignment <0.26 mm), the relative $TCF$ loss compared to the ideal alignment case is less than 1.4% ($TCF/TCF_0 > 98.6\%$). If the output surface of the CPC lens maintains full contact with the input surface of the rod lens (translational misalignment <0.5 mm), the $TCF/TCF_0$ ratio shows virtually no loss ($TCF/TCF_0 > 99.99\%$). When the diamond or CPC lens is rotated relative to the optical axis, introducing an angular misalignment within 10°, the $TCF/TCF_0$ remains over 98.2%. In practical bonding, manual alignment by visual inspection is sufficient to meet the above misalignment limits. With mechanical fixturing, translational misalignment can be controlled within 0.1 mm and angular misalignment within 1°, leading to even lower $TCF$ losses. The associated photon-count-rate reduction is proportional to $TCF$ and negligible, and the same holds for the impact on sensitivity. The assembly of the diamond probe is reproducible and satisfies the aforementioned misalignment constraints.

*3.2 Experiments results of diamond probes*

The experiment setup is described in detail in Fig. S3 in the Supplementary Material. The fluorescence intensity depended on the laser power obtained with various optical designs is shown in Fig. 4(a). The relationship between the fluorescence intensity and the laser power is nearly linear due to high linear correlation coefficient ($R^2 > 99\%$). The measured fluorescence data are nearly identical across five measurements, with a low standard error ($SE < 1.4\ \mu W$, relative $SE < 1\%$ and minimal deviation within the 95% confidence interval). The measured data is fitted to a linear module. The fitted slopes, shown in Table S2 in detail, reflect the FL collection level of various structures. Fig. 4(a) shows the FL intensity of five diamond probes. The FD-4 & optimized CPC lens (probe 1) exhibits a laser energy conversion efficiency of 0.4% (a 40 mW laser power generates 159.6 μW of fluorescence) and has the highest fluorescence collection level among the five structures. The FL collection capabilities of the FD-4 & optimized CPC, the cube diamond & optimized CPC (probe 2) increases by 37% and 1%, respectively, compared to the cube diamond & commercial CPC (probe 5, reference probe), while that of the cube diamond & tapered rod (probe 3) and the cube diamond & short CPC (probe 4) decreases by 32% and 53%, respectively. As a reference probe, the FL collection capability of probe 5 is just lower than probe 2 and probe 1. Therefore, there is no need to investigate the sensitivity levels of probe 3 and probe 4. The maximum laser power corresponds to the saturated optical power (40 mW) constrained by the PD's saturation photocurrent, which reflects the maximum collectable fluorescence power achievable in this experiment.

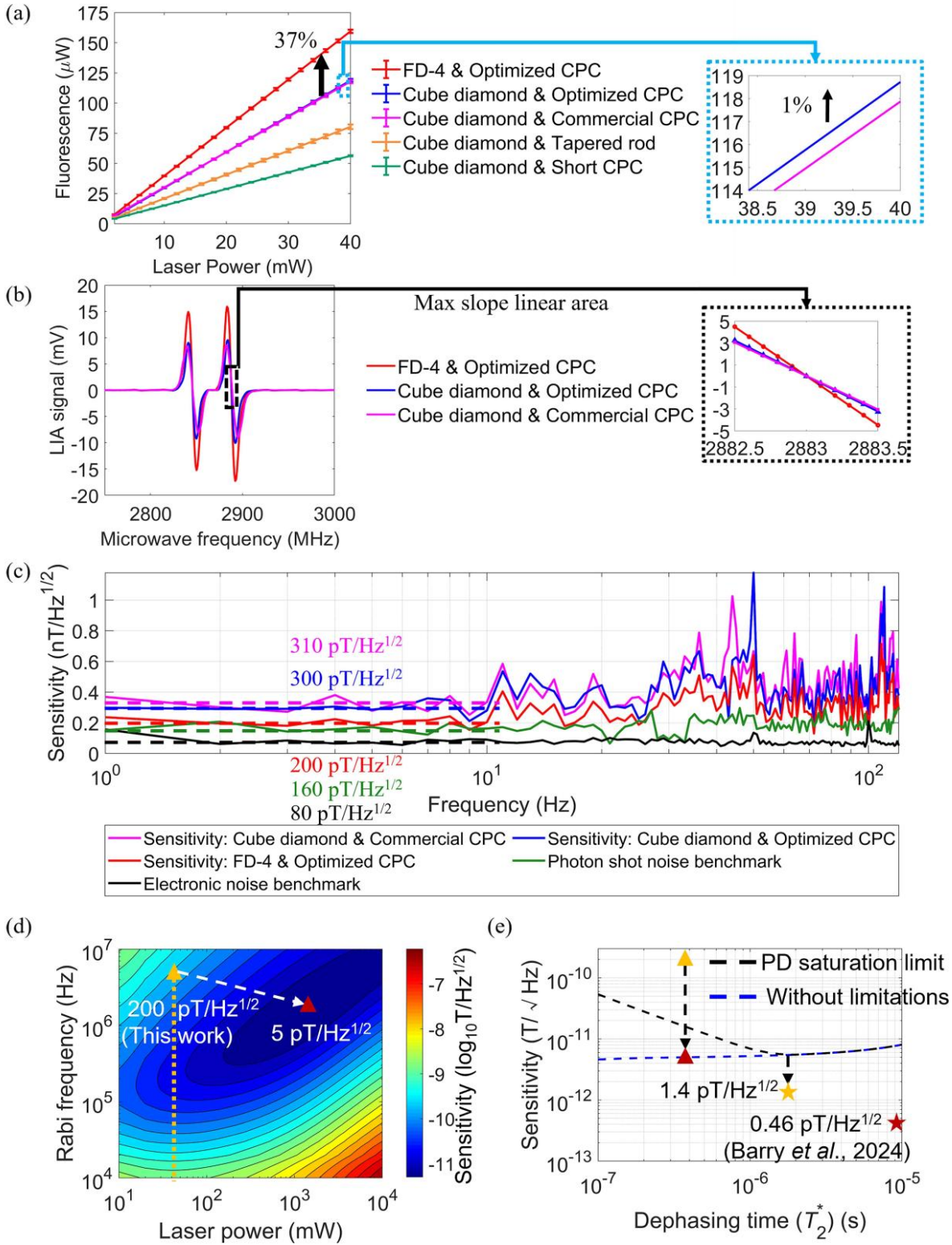

**Fig. 4.** Experiment results of diamond probes. (a) Error bar diagrams for different FL collection optical designs of collected fluorescence as a function of laser power variation. The fitted slopes reflect the FL collection level of various structures. (b) Demodulation signals of ODMR spectra obtained by three of the highest FL collection optical designs. The inset shows the maximum slope linear region. The maximum slope value reflects the magnetometer response to the external magnetic field. (c) Magnetic field sensitivity of such three diamond probes. The green and black lines represent the sensitivity levels

achievable by probe 1 when influenced solely by photon shot noise and electronic noise, respectively, without interference from other noise sources. (d) Potential pathways for enhancing the sensitivity of the diamond probe, which depend on both laser power and Rabi frequency. The sensitivity in the figure is the shot-noise-limited sensitivity. The orange dashed line indicates the maximum laser power (40 mW) achievable in this experiment, while the white dashed line represents the highest attainable sensitivity under fixed sample conditions (i.e., with a constant dephasing time). The yellow triangle indicates the sensitivity achieved in this study, while the red triangle represents the optimal sensitivity (5 pT/Hz$^{1/2}$) attainable under this dephasing time condition ($T_2^* = 350\ ns$). (e) Sensitivity as a function of dephasing time $T_2^*$. Each sensitivity value corresponds to the optimal combination of laser power and Rabi frequency for a given $T_2^*$. The black dashed line accounts for the impact of the PD saturation photocurrent on the collected fluorescence photon count, while the blue line does not consider this effect. The yellow pentagram represents the further achievable sensitivity through techniques such as hyperfine driving and double resonance based on CW-ODMR, whereas the red pentagram denotes the sub-pT level sensitivity attained using Ramsey sequences in previous studies (Barry et al., 2024) [27].

To demonstrate the performance of the three FL collection structures on the sensitivity of the diamond magnetometer, we employ the microwave frequency sweeping technique to obtain the ODMR spectra of such three probes. Moreover, we apply the MW frequency modulation technique to suppress low-frequency noise and enhance the signal-to-noise ratio (SNR) of the system. The modulation frequency is optimized as 4 kHz and the modulation depth is 8 MHz for the highest sensitivity. The laser power is set as the same when comparing the three probes and it is the saturation power of the probe 1. The MW power is set as its maximum to obtain the highest sensitivity. Fig. 4(b) shows the demodulated ODMR spectrum signals of three probes, which are obtained by the lock-in amplifier (LIA). Diamond sample was produced with a polished surface degree of (100). When a magnetic field is applied along the [100] crystallographic direction of the diamond, the four NV axes experience identical projections of the magnetic field. This configuration results in an ODMR spectrum featuring two distinct resonance peaks. Following lock-in demodulation, these peaks manifest as two linearly sensitive resonance regions centered at microwave frequencies of 2846 MHz and 2888 MHz, respectively. The magnetic field and temperature can be derived from the following equations [8][17].

$$B_{ext} = \frac{B_{NV}}{\cos(\theta)} = \frac{\sqrt{3}}{2\gamma_e}(f_+ - f_-) \tag{4}$$

$$\Delta T_{probe} = \frac{1}{\beta_T}\left[\frac{(f_+ + f_-)}{2} - D_{gs}\right] \tag{5}$$

Here, refers to the external magnetic field strength $B_{ext}$, which forms an angle $\theta$ with the internal magnetic field sensed along the NV axis $B_{NV}$. The internal field $B_{NV}$ can be resolved from the ODMR resonance frequencies, where $f_+$ corresponds to the resonance peak of the spin sublevel $|m_s = +1>$ (2888 MHz in this experiment) and $f_-$ to that of the sublevel $|m_s = -1>$ (2846 MHz in this experiment). $\gamma_e$ is the electron gyromagnetic ratio, taken as 2.8 MHz/G. $\Delta T_{probe}$ represents the temperature change of the probe relative to room temperature, and $\beta_T$ is the temperature coefficient, with a value of about –74.2 kHz/K. $D_{gs}$ denotes the zero-field splitting at room temperature (300 K), which is 2867 MHz and varies with sample properties and external conditions [35]. From the equations above, the sensed magnetic field and temperature under the experimental conditions of this study are derived as 13 G and 300 K (laboratory temperature: 298 K, temperature rise: 2 K), respectively. These specific frequencies correspond to the energy level differences arising from the Zeeman splitting between $|m_s = 0>$ state and the $|m_s = \pm 1>$ states. These two resonance peaks occur at the points of maximum slope within the demodulated ODMR spectrum, indicated by the areas highlighted within the black boxes in the Fig. 4(b). To determine the scalar factors for the three probes, frequency sweeps are conducted within a 1-MHz range centered around each maximum slope region, utilizing a step size of 0.1 MHz. The resulting scalar factors for probe 1, probe 2, and probe 3 are determined to be 8.91 mV/MHz, 6.47 mV/MHz, and 6.07 mV/MHz, respectively.

Having obtained the scalar factors at the maximum slope points of the demodulated ODMR spectrum, the magnetic field sensitivity of each probe can then be determined by measuring the noise floor of the lock-in output signal. The estimated sensitivity of the magnetometers can be derived from the power-spectral-density (PSD) plot [36], as shown in Fig. 4(c). The plot is obtained by dividing by two coefficients ($\gamma_e$ and $k$), where $\gamma_e = 2.8$ MHz/G is the gyromagnetic ratio of electron spin, $k = \max(|dV/df|)$ is the scalar factor. The PSD plot is measured by the LIA. Fig. 4(c) presents the sensitivity levels of three probes alongside the electronic noise benchmark. The sensitivities of probe 1 to probe 3 reach 200 pT/Hz$^{1/2}$, 300 pT/Hz$^{1/2}$, and 310 pT/Hz$^{1/2}$, respectively, demonstrating that probe 1, associated with superior FL collection capability, exhibits the highest sensitivity level. Due to the magnetometer's frequency-dependent response characteristics and the bandwidth setting (bandwidth: 200 Hz) of the LIA filter, the sensitivity values vary across different frequency bands. The reported sensitivity values represent the average of the PSD measurements obtained within the 1-10 Hz frequency range. Furthermore, the data were acquired over a 20-second sampling duration at a sampling rate of 3598 SPS. The photon shot noise benchmark, measured as 160 pT/Hz$^{1/2}$, indicates the sensitivity level of the diamond determined by photon shot noise. This noise level corresponds to the saturation laser power of the PD at 40 mW, which reflects the attainable sensitivity under photon shot noise limitation. Under the same scaling factor, a lower photon shot noise would result in a higher sensitivity level. The electronic noise benchmark, measured at 80 pT/Hz$^{1/2}$, represents the fundamental sensitivity limit achievable when the combined electronic noise floor of the LIA, transimpedance amplifier (TIA) and PD dominates the overall system noise. This value simultaneously reflects the upper sensitivity limit constrained by the demodulation circuitry, the transimpedance circuitry and the PD properties (e.g. dark current). A detailed noise analysis is provided in Supplementary Material 7. It is shown from Fig. 4(c) and Fig. S5 that the overall noise level of the magnetometer is close to the photon shot noise, while a more significant difference exists between photon shot noise and electronic noise. A similar trend is observed in the corresponding sensitivity, indicating that the sensitivity of the diamond probe is primarily limited by photon shot noise. Reducing electronic noise from components such as the LIA, TIA, and PD can lower overall noise and improve sensitivity, but the resulting enhancement is less significant compared to that achieved by reducing photon shot noise. Therefore, the magnetometer can be optimized to further improve the measurement sensitivity by noise suppression.

The measured actual sensitivity in this study is 200 pT/Hz$^{1/2}$. Fig. 4(d) and (e) discuss how higher sensitivity can be achieved through improved technical approaches, such as replacing the PD with one that has a higher saturation photocurrent, increasing the laser power, using samples with longer dephasing time, and applying techniques including hyperfine driving, double resonance, and Ramsey sequences. This sensitivity calculation is purely based on theoretical simulation and estimation, primarily focusing on potential pathways to achieve superior sensitivity based on the probe design in this study. Due to the saturation photocurrent limit (90 μA) of the PD (responsivity: 0.55 A/W; saturation laser power: 40 mW), fluorescence photons excited by higher laser power cannot be effectively converted into electrical signals, thereby preventing further improvement in sensitivity. Detailed parameters and analytical calculations regarding the sensitivity estimation are provided in the Supplementary Material 6. Fig. 4(d) illustrates the relationship between sensitivity, laser power, and Rabi frequency under fixed sample conditions (with a constant dephasing time, $T_2^* = 350$ ns), reflecting the potential sensitivity enhancement capability of this sample. The measured dephasing time ($T_2^*$) of the sample used in this study is 350 ns, and its value is used for the sensitivity estimation in Fig. 4(d) and (e). The sensitivity relationship is derived from the CW-ODMR expressions (Equations (S11), (S12), and (S13)). Under the photon shot noise limit, sensitivity is described by Equation $\eta \approx P_F \frac{h}{g\mu_B} \frac{\Delta\nu}{C\sqrt{R_{fl}}}$ (1) and Equation (S11). In these equations, the linewidth $\Delta\nu$ and contrast $C$ depend on the optical saturation parameter $s$ and the Rabi frequency $\Omega_R$, with $s$ itself being a function of the input laser power $P$. Consequently, the shot-noise-limited sensitivity can be expressed in terms of both laser power and Rabi frequency, consistent with the description in reference [15]. This formulation combines the detected fluorescence photon rate, proportional to $R_{fl}$, and the photon shot noise, proportional to $\sqrt{R_{fl}}$. Each sensitivity value plotted in Fig. 4(d) corresponds to the photon-shot-noise-limited sensitivity at a specified laser power and Rabi frequency. The red triangle indicates the actual sensitivity level measured for probe 1, while the yellow pentagram marks the potentially achievable maximum sensitivity after optimization, with a value of 5 pT/Hz$^{1/2}$. The maximum laser power employed in this study is 40 mW. However, achieving the full sensitivity potential of this sample requires a laser power of 1.43 W, which would generate a photocurrent of 0.88 mA—placing significantly higher demands on the saturation photocurrent tolerance of the PD. Under the sample condition, the optimal achievable sensitivity (5 pT/Hz$^{1/2}$) was primarily estimated based on the theoretical limit set by photon shot noise. In this study, however, the actual sensitivity is primarily constrained by the insufficient saturation photocurrent of the PD, which limits the conversion of a larger number of fluorescence photons into photocurrent, according to Equation $\eta \approx P_F \frac{h}{g\mu_B} \frac{\Delta\nu}{C\sqrt{R_{fl}}}$ (1).

Fig. 4(e) shows the sensitivity levels for different samples (with varying $T_2^*$), where the laser power and Rabi frequency are optimized for each dephasing time. The figure reflects the probe sensitivity that could be attained if a higher-quality sample were employed in this experiment. However, fabricating samples with long dephasing time involves complex processes and higher costs. Therefore, this estimation only serves as a projection of the expected performance under the probe design conditions adopted in this study. The blue dashed curve accounts for the NV concentration (inversely correlated) under different dephasing time conditions, which affects the number of fluorescence photons generated (positively correlated with the concentration). When the saturation photocurrent of the PD is not considered—i.e., assuming all fluorescence emitted by the diamond is collected by the PD—the optimal sensitivity is strongly influenced by the photon count rate, leading to a trend where sensitivity degrades as $T_2^*$ increases. However, when the limitation imposed by the PD's saturation photocurrent is considered, the high fluorescence yield from samples with short $T_2^*$ (higher NV concentration) no longer provides an advantage. Instead, the sensitivity exhibits a minimum point as $T_2^*$ varies, as indicated by the black dashed curve in the figure. Furthermore, a PD with a higher saturation photocurrent generally has a higher noise equivalent power (NEP), and the detection SNR becomes limited by the noise introduced by the PD. For samples with longer $T_2^*$, the photocurrent generated by the fluorescence under optimal laser power is below the PD's saturation threshold. On this condition, sensitivity is primarily governed by the available fluorescence photon count rate. Although higher NV concentration can improve sensitivity, it also demands higher laser and microwave power. Thus, samples with longer $T_2^*$ require lower laser and microwave power, contributing to reduced overall system power consumption. All sensitivity estimations shown in the figure are based on CW-ODMR detection. Under conditions considering PD saturation limitations, the optimal sensitivity reaches 5.5 pT/Hz$^{1/2}$. When hyperfine driving and double resonance techniques are incorporated, sensitivity can be further improved to 1.4 pT/Hz$^{1/2}$. The red pentagram in the figure represents the results (Barry et al., 2024) [27] where a Ramsey sequence was employed. The study achieves a measurement sensitivity of 460 fT/Hz$^{1/2}$, outperforming CW-ODMR and highlighting a potential direction for further improvement in this work. The sample used in that study has a dephasing time of 8.7 $\mu$s, which is not limited by the PD saturation photocurrent, and its sensitivity is significantly influenced by the fluorescence photon count rate. The fluorescence collection efficiency enhancement technique developed in this work (37% enhancement) holds potential for further increasing the sensitivity. The improvement in fluorescence collection efficiency and sensitivity achieved through this optical design optimization not only meets the key metrics required for biomagnetic measurements using medical endoscopes, but also demonstrates the potential application prospects of millimeter-scale rigid diamond probes.

## 4. Conclusion

In summary, we have developed a millimeter-scale diamond magnetometer probe that achieves 200 pT/Hz$^{1/2}$ sensitivity through optical designs. This study addresses the critical trade-off between miniaturization and sensitivity in diamond magnetometry by developing an integrated optical architecture. By synergistically combining a frustum diamond with 4 inclined planes (FD-4) with a miniaturized compound parabolic concentrator (CPC) lens, we achieved a 37% enhancement in fluorescence collection efficiency within a 4 mm diameter probe dimension, which is a size reduction surpassing current state-of-the-art optical collectors (e.g., TIR lenses, commercial CPC lenses). The optimized CPC design (3 mm diameter) demonstrates superior photon transmission characteristics over commercial alternatives, while the FD-4 maximizes total emitted fluorescence efficiency, enabling a higher signal-to-noise ratio and a higher sensitivity at 8 MHz ODMR linewidth.

The synergistic integration establishes a paradigm in endoscopic quantum sensing. This architecture achieves a critical balance between miniaturization and sensitivity. This work enables transformative applications across real-time intraoperative monitoring and industrial diagnostics. Future work will focus on further validation of this endoscopic diamond probe in both endoscopic scenarios and in situ environments with the ultimate goal of enabling its successful application in the field of medical endoscopic magnetometry. Thermal analysis and robustness analysis under mechanical stress will also be investigated in detail for biomagnetic measurements in endoscopic scenarios. The sensitivity enhancement strategy proposed in this work offers a potential approach for quantum-based biological endoscopy, thereby establishing a bridge between quantum precision measurement and practical diagnosis.

**ACKNOWLEDGMENTS**

This work was supported by the National Key R&D Program of China (Grant No.

2024YFC3606900); the Projects of the National Natural Science Foundation of China under Grant Nos. 62173020, 62473058, 61773046, 61673041; the Key Area Research and Development Program of Guangdong Province under Grant2021B0101410005.

# Supplementary material: Millimeter-scale rigid diamond probe for high sensitivity endoscopic-magnetometry applications

## 1. Analysis of diamond emitted fluorescence

The diamond structure is obtained by inclined cutting and polishing the four sides of the cube diamond, forming a frustum diamond structure. The diamond structures include inclined cutting on two parallel sides and all four sides to form a frustum diamond with 2 inclined planes (FD-2) and a frustum diamond with 4 inclined planes (FD-4). When the inclination angle reaches its limit, it forms a triangular prism diamond (TPD) and a pyramid diamond (PYD), as shown in Fig. S5(a). Different diamond structures exhibit varying efficiency ($\eta_e$) of emitted fluorescence (FL) when contacted by different surrounding media. This comparative result is displayed in Fig. S5(b) and (c). Due to the loss of NV centers during the cutting process of a cube diamond, the number of NV centers decreases, resulting in a reduction in the total FL generated inside the diamond. Therefore, the total emitted fluorescence ($TEF$) from the diamond is defined as the product of the volume ratio ($V_s/V_c$) and the emitted efficiency ($\eta_e$, $TEF = V_s/V_c \times \eta_e$). The total FL photons generated by the frustum diamond ($R_s$) are described by the following formula:

$$R_s = R_c V_s / V_c \quad (S1)$$

where $V_c$ is the volume of the cube diamond (0.125 mm³ in this study), $V_s$ is the volume of the diamond under frustum structures, $R_c$ is the total FL photons generated in the cube diamond. Equation $R_s = R_c V_s/V_c$ (S1) is based on the sample that NV centers are approximately uniformly distributed within the diamond. In this case, $TEF$ and $\eta_e$ represent the proportion of the total FL photons generated inside the cube diamond and the frustum diamond respectively. On this condition, the total collected fluorescence ($TCF$) and total collected FL photons ($R_{fl}$) by the photodetector (PD) are defined as follows:

$$TCF = TEF \times \eta_c = \eta_e \eta_c V_s / V_c \quad (S2)$$
$$R_{fl} = R_c \times TCF \quad (S3)$$

where $\eta_c$ represents the efficiency of collected emitted FL from the diamond. $TCF$ represents the proportion of the FL collected by the PD relative to the total FL generated inside the (0.5 mm)³ cube diamond. It is proportional to the total collected FL photons ($R_{fl}$) by the PD ($TCF \propto R_{fl}$), thereby determining the sensitivity of the diamond magnetometer.

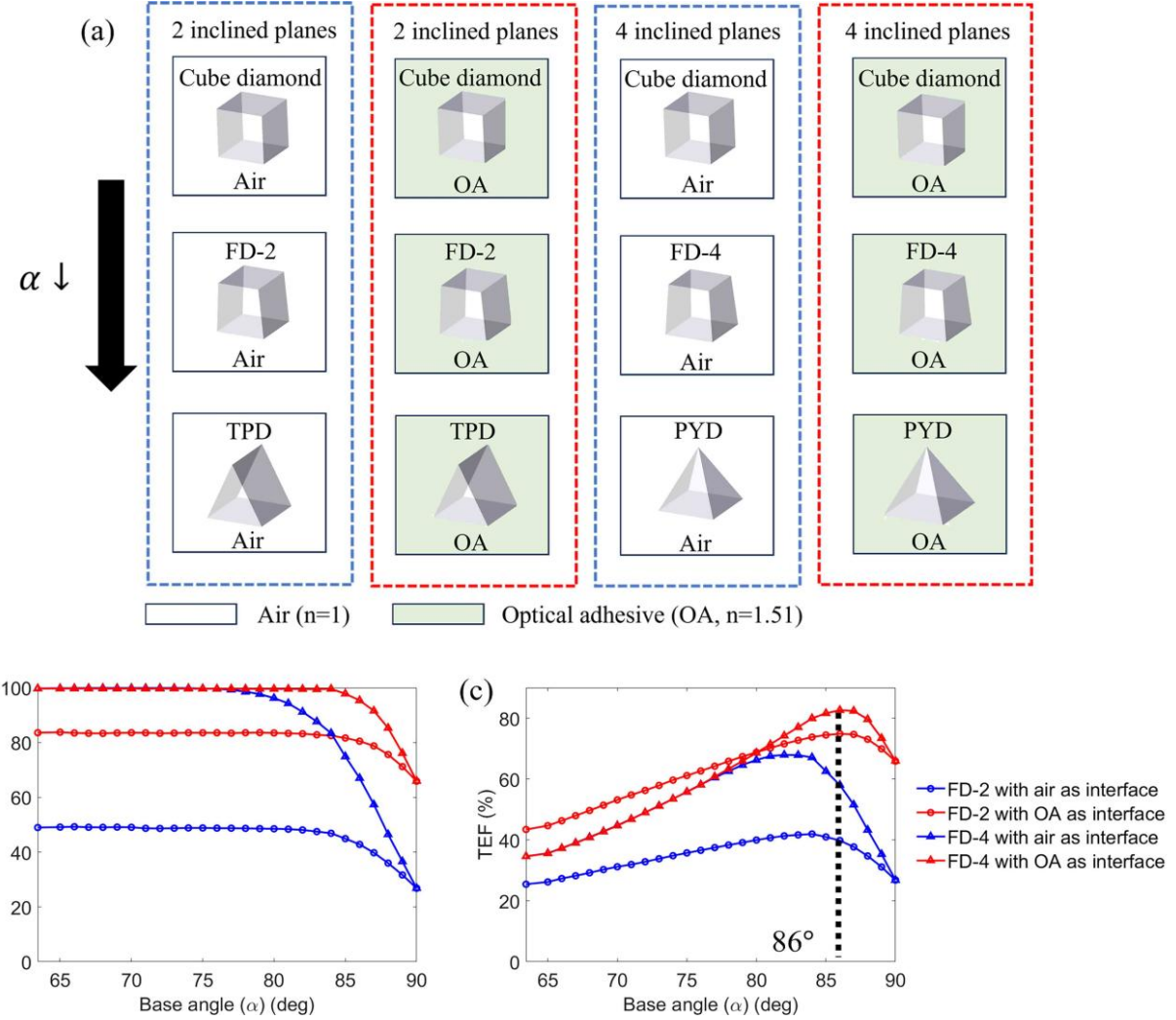

**Fig. S5.** (a) The diamond structure variations. These illustrate the changes in diamond structures with different cutting methods and surrounding interfaces. (b) The relationship between total emitted efficiency ($\eta_e$) and the base angle ($\alpha$) of the inclined surface. (c) The relationship between total emitted fluorescence ($TEF$) and the base angle ($\alpha$) of the inclined surface. With the base angle ($\alpha$) decreases, the total emitted efficiency increases while the diamond volume decreases, which reflects the number of NV centers (FL sources) decreases.

When $\alpha$ increases, $V_s$ increases but $\eta_e$ decreases, resulting in a peak in Fig. S5(c). When surrounded by air, only 27% of the FL escapes from a cube diamond. The optical adhesive layer ($n = 1.51$) is a diamond and lens transition layer. It can enable FL to exit the diamond-air interface at a wider range of incidence angles, resulting in a development of 40% in the emitting efficiency. When optimizing the diamond structure to construct a frustum with two or four inclined planes, the emitting efficiency can approach nearly 100% when $\alpha$ is reduced to below 75°. However, constructing frustum results in losing internal FL generated in the diamond. When the base angle is between 80° and 87°, the $TEF$ can reach its maximum. Therefore, changing the diamond structure to a frustum structure with 4 inclined planes and surrounding it with an adhesive layer can effectively enhance the emitted FL of the diamond.

## 2. Optical designs

The two-dimensional frustum diamond structure is shown in Fig. S6(a). The variation of the front surface line $l_f$ with the base angle $\alpha$ can be expressed as below:

$$l_f(\text{mm}) = 0.5 - 1/\tan\alpha \tag{S4}$$

the frustum diamond structure is determined by the cutting slope base angle $\alpha$ and the cutting of the side surfaces.

The CPC lens exhibits a rotationally symmetric structure, as is shown in Fig. S6(b). The input radius ($r_{in}$) determines the dual parabolic surface of the CPC lens and the maximum divergence angle of FL inside the lens ($\theta_{lm}$), which is defined as follows:

$$z(r) = \left[r_{in}\cos\theta_{lm}(2+\sin\theta_{lm}) - r\cos\theta_{lm}\sin\theta_{lm} - 2\sqrt{r_{in}(1+\sin\theta_{lm})}\sqrt{r_{in}-r\sin\theta_{lm}}\right]/\sin^2(\theta_{lm}) \tag{S5}$$

where $r$ is the distance from the z-axis and $z(r)$ is the surface displacement from the vertex along the z-direction, assuming the optical axis lies along the z-direction. The CPC output radius $r_{out}$ and length $L$ determine the size of the CPC and are also determined by the input radius $r_{in}$ and the maximum divergence angle of FL inside the lens ($\theta_{lm}$). According to the refraction theory, the FL ray exhibits a larger divergence angle when it enters the air, which is referred to as the maximum divergence angle of FL in the air ($\theta_{am}$). The relevant definitions are as follows:

$$r_{out} = r_{in}/\sin\theta_{lm} \tag{S6}$$
$$L = (r_{in} + r_{out})/\tan\theta_{lm} \tag{S7}$$
$$\theta_{am} = \arcsin(n_{air}\sin\theta_{lm}/n_{lens}) \tag{S8}$$

where $n_{air}$ represents the refractive index of air and $n_{lens}$ represents the refractive index of the lens.

The schematic diagram of a tapered rod is shown in Fig. S6(c). The tapered rod is used as the collimating lens and the FL collecting lens. The parameters used in the tapered rod lens is the same as the CPC lens. Its facial formula can be described as follows:

$$z(r) = (1+\sin\theta_{lm})(r-r_{in})/[(1-\sin\theta_{lm})\tan\theta_{lm}] \tag{S9}$$

however, the parameters can be set alone, which is different from the CPC lens.

The coated aspheric lens is illustrated in Fig. S6(d). During optimization, the aspheric lens is typically optimized and simplified to a parabolic lens, determined solely by its length ($L_a$), maximum radius ($R_a$), curvature radius ($R_c$), and curvature coefficient ($k$). When $k$ is certain, $R_c$ is determined by $L_a$ and $R_a$. For the parabolic face, $k = -1$. Equation $z(r) = r^2/[R_c(1+\sqrt{1-(1+k)\frac{r^2}{R_c^2}})]$(S10) illustrates the surface profile formula for the aspheric surface and higher-order terms are neglected:

$$z(r) = r^2/[R_c(1+\sqrt{1-(1+k)\frac{r^2}{R_c^2}})] \tag{S10}$$

where $r$ is the distance from the z-axis and $z(r)$ is the surface displacement from the vertex along the z-direction, assuming the optical axis lies along the z-direction. $R_c$ and $k$ determine the surface profile structure of the parabolic surface while $L_a$ and $R_a$ determine the size of the parabolic surface. Maximum radius ($R_a$) is limited by the input radius of the CPC lens. Therefore, the surface profile structure is only determined by $L_a$. To ensure that the diamond is fully embedded, the minimum $L_a$ is 0.8 mm.

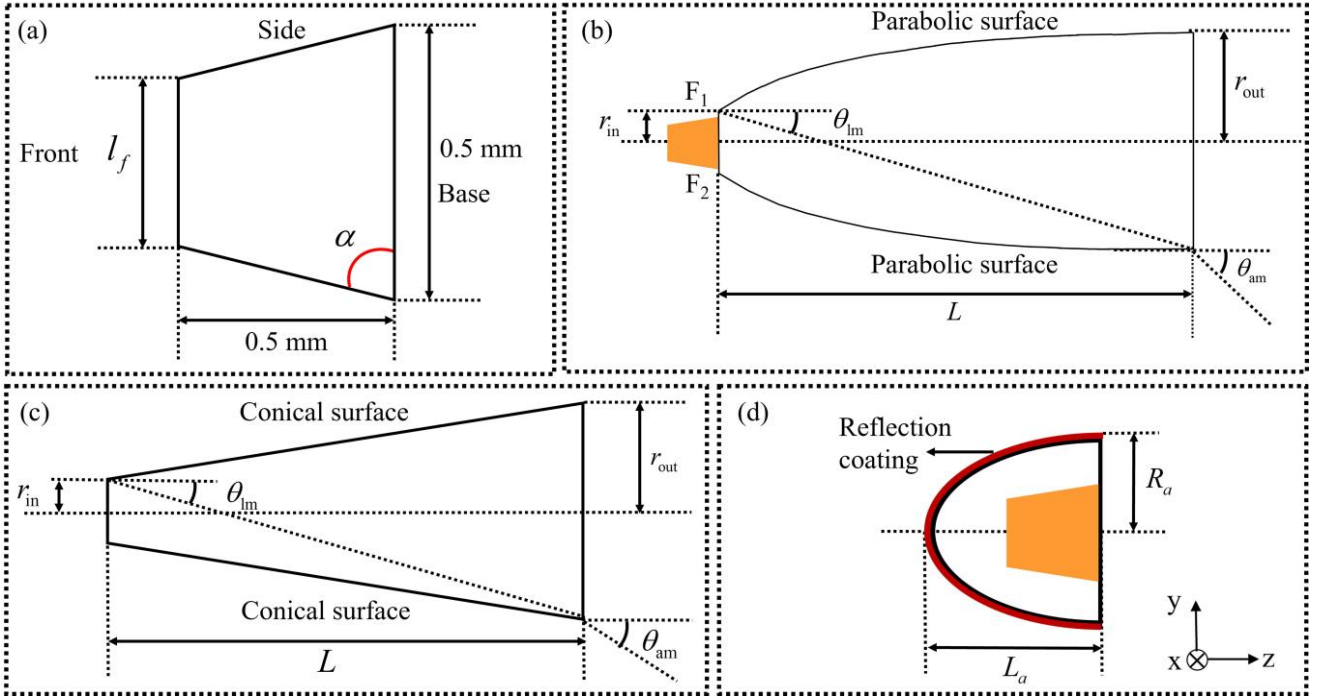

**Fig. S6.** (a) Frustum diamond structure. The length of the base face is 0.5 mm, and the side surface vertical length is 0.5 mm. The length of the front face ($l_f$) decreases when reducing the cutting slope base angle ($\alpha$). (b) CPC lens structure. It comprises two parabolic surfaces with focal points F1 and F2. The divergence angle of the FL inside the lens ($\theta_l$) decreases when reflected from the parabolic surface. The final divergence angle is smaller than the maximum divergence angle of the FL inside the lens ($\theta_{lm}$). When the FL is emitted into the air, the divergence angle in the air ($\theta_a$) is smaller than the maximum divergence angle of the FL in the air ($\theta_{am}$). (c) Schematic diagram of the tapered rod lens. (d) Coated aspheric lens, with the diamond embedded in and with the reflection coating on the outer surface of the lens.

## 3. Experiment set up

The simplified structure of the experiment system is shown in Fig. S7. The 532 nm laser is irradiated on the diamond to excite NV centers. One pathway of the laser excites the diamond NV centers while the other is the reference signal for differential fluorescence signal analysis. The laser differential technique is beneficial for suppressing laser noise. The two transimpedance amplifiers (TIA) output signal are made as the same for noise suppression. The fluorescence emitted from the diamond is collected by the CPC lens and the rod lens, and detected by a photodiode. A dichroic prism and a long-pass filter with 665 nm cut-off wavelength are placed after the rod lens to filter 532 nm laser light. The dichroic prism can also reflect the 532 nm laser to excite the NV centers. The PD converts the FL into a current signal (gain: 0.55 A/W; short circuit current: 90 μA). Based on the fluorescence signal characteristics, a TIA is designed to effectively convert the photocurrent in the photodiode into voltage (gain: 5.1 kV/A). Its electrical voltage signal is connected with a lock-in amplifier (LIA) to facilitate the subsequent signal processing.

Moreover, the microwave (MW) signal is generated from a MW source and sent through a switch to a high-power amplifier and finally delivered by an EMC near-field probe. The MW signal is modulated by modulation signal generated by a signal generator. The microwave frequency modulation can be expressed as $f_{MW} = f_c + f_{dev}\cos2\pi f_{mod}t$, where $f_c$ represents the microwave central frequency, $f_{mod}$ represents the modulation frequency, and $f_{dev}$ represents the modulation depth. The optimized modulation frequency and depth are decided by the sensitivity. The fluorescence exhibits the same modulation frequency when modulating the microwave frequency. The reference signal is generated by the same source and exhibits the same frequency. The FL signal and the reference signal are both transmitted to a lock-in amplifier for demodulation. Therefore, the weak magnetic signal can be detected with a high signal-to-noise ratio (SNR).

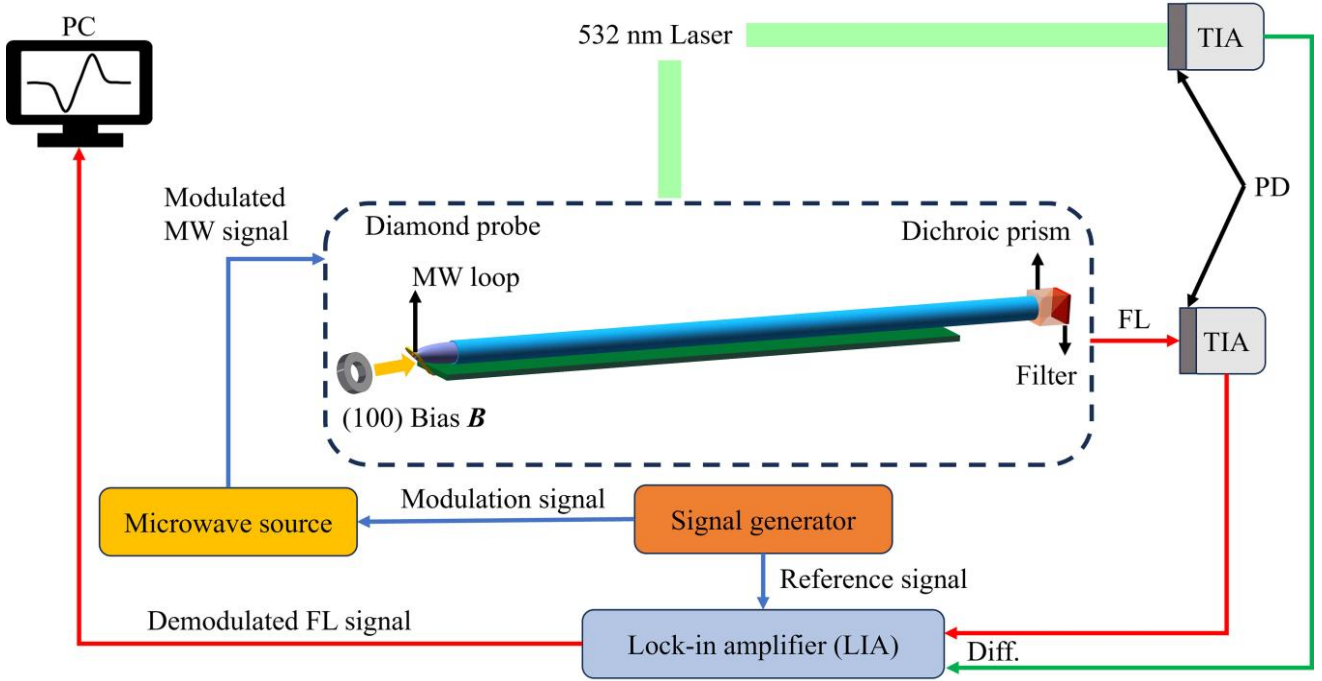

**Fig. S7.** The schematic diagram of the experiment system.

## 4. Optimized parameters

The optimized parameters of all optical designs are shown in Table S2. The results are based on the simulation results in Fig. 3.

**Table S2**
Optimized parameters of all designs.

| Design | Optimized parameters | | | |
|---|---|---|---|---|
| Diamond | Inclined plane | $\alpha$ (deg) | | $l_f$ (mm) |
| | 4 planes | 86.0 | | 0.43 |
| CPC | $r_{out}$ (mm) | $\theta_{am}$ (deg) | $r_{in}$ (mm) | $\theta_{lm}$ (deg) |
| ($n = 1.47$) | 1.50 | 30 | 0.51 | 19.9 |
| Tapered rod | $r_{out}$ (mm) | $\theta_{am}$ (deg) | $r_{in}$ (mm) | $\theta_{lm}$ (deg) |
| ($n = 1.90$) | 1.70 | 31.5 | 0.47 | 15.6 |
| Coated aspheric lens | Coating | $L_a$ (mm) | $R_a$ (mm) | $k$ $R_c$(mm) |
| | $Ta_2O_5$-$Al_2O_3$ | 0.80 | 0.50 | -1.00   0.16 |

## 5. Fluorescence collection by laser power

The slopes of different FL collection optical designs are shown in **Table S3**. Standard errors of the slopes are based on 95% confidence interval of normal distribution. The slopes are based on the FL intensity experiment results in Fig. 4(a).

**Table S3**
Slopes of various FL collection optical designs in Fig. 4(a).

| FL collection optical design | | Fluorescence/Laser power (×10⁻³) | |
|---|---|---|---|
| Diamond | Lens | Average | Standard error (95% confidence interval) |
| Cube diamond | Short CPC | 1.37 | 1.57×10⁻⁴ |
| | Tapered rod | 1.98 | 2.90×10⁻³ |
| | Commercial CPC | 2.92 | 1.42×10⁻³ |
| | Optimized CPC | 2.96 | 1.31×10⁻³ |
| FD-4 | Optimized CPC | 3.99 | 1.56×10⁻³ |

## 6. Sensitivity estimation of the diamond probe

Since this study employs continuous-wave ODMR (CW-ODMR) to reflect the capability in magnetic field measurement and fluorescence collection, the sensitivity estimation formula for CW-ODMR is selected as the theoretical basis for sensitivity estimation of the proposed diamond probe, as expressed below:

$$\eta_{\text{CW-ODMR}} \approx P_F \frac{h}{g\mu_B} \frac{\Delta\nu}{C\sqrt{R_{\text{fl}}}} \quad (S11)$$

where $\Delta\nu$ is the linewidth of the ODMR spectrum, $C$ is the signal contrast, $R_{\text{fl}}$ is the collected photons by the PD, $P_F$ is a factor related to the line shape (for Lorentzian line profile used in the present work, $P_F = 0.77$), $h = 6.626 \times 10^{-34}$ Js is the Planck constant, $g = 2$ is the electron g-factor and $\mu_B = 9.274 \times 10^{-24}$ J/T is the Bohr magneton. The formulas for calculating linewidth and contrast are given as follows:

$$C = \frac{1}{2} \frac{(a-b)\Gamma_p}{(a+b)\Gamma_1 + a\Gamma_p} \frac{\Omega_R^2}{\Omega_R^2 + \Gamma_2(2\Gamma_1 + \Gamma_p)} \quad (S12)$$

$$\Delta\nu = \frac{1}{2\pi}\sqrt{\Gamma_2^2 + \frac{\Omega_R^2}{\Omega_R^2 + \Gamma_2(2\Gamma_1 + \Gamma_p)}} \quad (S13)$$

where $\Gamma_1 = 1/T_1$ ($T_1 = 0.8$ ms) is the spin-lattice relaxation rate and $\Gamma_2^* = 1/T_2^*$ ($T_2^* = 350$ ns in this work) is the spin-spin-relaxation rate ($\Gamma_2 = \Gamma_2^* + \Gamma_c$). The optical pumping rate is defined as $\Gamma_c = \Gamma_c^\infty s/(s+1)$ ($\Gamma_c^\infty = 8 \times 10^7$ s⁻¹) and the optical repolarization rate is $\Gamma_p = \Gamma_p^\infty s/(s+1)$ ($\Gamma_p^\infty = 5 \times 10^6$ s⁻¹). The optical saturation coefficient $s$ is defined as $s = P/P_{\text{sat}}$ ($P_{\text{sat}} = 34$ W, $P$ is the laser power). For the contrast calculation, the coefficient $a$ is related to the maximum contrast $C_{max}$ ($a = 1/(1 - 2C_{max})$) and the coefficient $b = 1$. As is shown in **Fig. S8**, since a [100]-oriented magnetic field is applied to the diamond with [100] crystallographic orientation in this experiment, the magnetic field projections along all four NV axes are identical, causing the four resonance peaks in the ODMR spectrum to overlap (peak a). The measured contrast is four times the estimated contrast for a single peak. When the magnetic field is applied along the [111] crystallographic direction, the projections of the field along three of the NV axes are identical (peak b₂), each being one-third of the external field magnitude, while the remaining NV axis aligns parallel to the field and experiences the full external field (peak b₁). This results in an ODMR spectrum featuring four distinct peaks (with three peaks overlapping, the contrast is three times that of the other peak), as shown in **Fig. S8**(b). In this experiment, the magnetic field is applied along the [100] crystallographic direction to achieve higher ODMR contrast. The maximum measured contrast is 0.1 ($C_{max} = 0.1$), which is substituted into the sensitivity estimation calculation here.

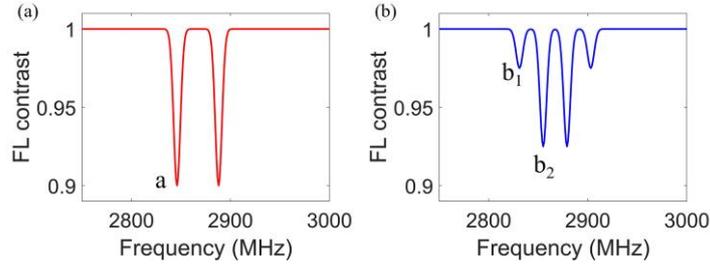

**Fig. S8.** ODMR spectrums under magnetic fields applied along different crystallographic orientations. (a) Magnetic field oriented along the [100] direction. (b) Magnetic field oriented along the [111] direction.

In the sensitivity estimation, the maximum laser power is limited to 40 mW due to the saturation photocurrent constraint of the PD (gain: 0.55 A/W), generating 159.6 µW of fluorescence photons. Based on this, it is derived that 1 mW of laser power produces 3.99 µW of fluorescence photons, equivalent to a photocurrent of 7.25 µA. This corresponds to an estimated photon count rate of $R_{fl} = I_{pd}/q$, ($q$ is the electron charge). The Rabi frequency ($\Omega_R$) in the experiment is 4.17 MHz (pi-pulse time is 120 ns). Based on the analysis and calculation of the aforementioned formulas and parameters, the relationship between sensitivity, laser power, and Rabi frequency is shown in Fig. 4(d), while the dependence of sensitivity on $T_2^*$ is presented in Fig. 4(e). For each value of $T_2^*$, both the laser power and Rabi frequency are optimized to their respective optimal values. In the analysis and discussion of Fig. 4(e), this study assumes an inverse relationship between $T_2^*$ and NV concentration ([NV]) (i.e., $1/T_2^* \approx \beta n_{NV} + \beta_0$, $\beta$ and $\beta_0$ are the linear coefficients). However, if the conversion efficiency from nitrogen atoms [N] to nitrogen-vacancy centers [NV] is low—resulting in a low [NV] concentration but a high [P1] center concentration—samples with shorter $T_2^*$ may also exhibit reduced [NV] density. Therefore, the relationship between $T_2^*$ and [NV] needs further research. This study assumes a proportional correlation between NV concentration and the fluorescence photon count rate. (i.e. $R_{fl} = R_{fl0} n_c/n_0$, $R_{fl0}$ is the photon count rate (1 mW laser power) in this work, $n_c$ is the [NV] related to the given $T_2^*$, and $n_0$ is the [NV] in this work). During the estimation of sensitivity as a function of $T_2^*$, all other parameters are held constant, with only the effects of NV concentration and photon count taken into consideration.

## 7. Noise analysis

Noise in the diamond magnetometer is co-dominated by optical and electronic sources according to Ref. [3]. When optical noise dominates, suppressing electronic noise yields limited improvement in magnetometer sensitivity, and vice versa. Fig. S5 presents a detailed noise analysis. The red curve represents the overall noise level determining the sensitivity of the diamond sensor, which is influenced collectively by photon shot noise, electronic shot noise, ambient magnetic fields, and other noise sources. Photon shot noise corresponds to the noise level observed when a 40 mW laser is applied to the diamond, with the PD in saturation and no microwave excitation applied. This noise level includes contributions from the PD, TIA, and LIA. The black curve depicts the noise output from the LIA alone, the combined LIA and TIA output, and the total noise including the PD. The measured photon and electronic noise floors (1–10 Hz) are 30.7 nV/Hz$^{1/2}$, 13.9 nV/Hz$^{1/2}$, 11.3 nV/Hz$^{1/2}$, and 3.5 nV/Hz$^{1/2}$, respectively. The results indicate that photon shot noise is the predominant factor limiting the sensitivity of the diamond sensor. While improving electronic noise can reduce the overall noise level and enhance the sensitivity of the diamond sensor, its effect is less pronounced compared to reducing photon shot noise. When the overall noise of the diamond system is primarily limited by the lock-in amplifier (the noise level is comparable to that of the LIA), a sensitivity of 15.7 pT/√Hz can be achieved. To reach the estimated sensitivity target of 5 pT/√Hz, it is necessary to enhance the optical signal intensity and employ a photodetector with a higher saturation photocurrent.

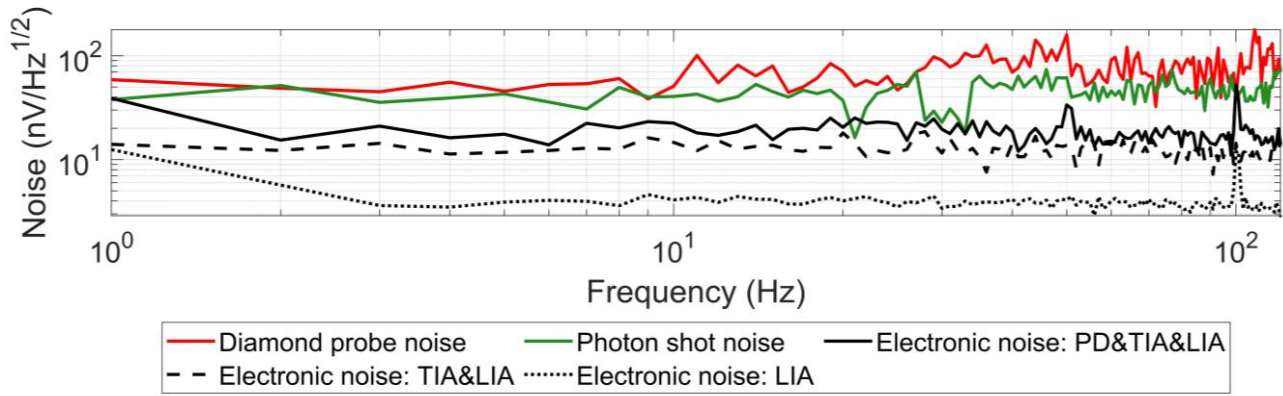

**Fig. S9.** Noise analysis affecting the sensitivity of diamond magnetometers. The red line represents the probe noise level that determines the sensitivity. The black line indicates the noise contributions from the lock-in amplifier (LIA), transimpedance amplifier (TIA), and photodiode (PD), while the green line represents the noise level when a 40 mW laser is applied in addition to the aforementioned electronic components.

## 8. MW loop

Fig. S6 shows the parameter characteristics of the microwave loop. The magnetic field generated by the applied microwave loop is sufficiently uniform across the diamond region, with a uniformity exceeding 95% over a 0.5 × 0.5 mm$^2$ area. The antenna impedance is matched to 50 Ω. The simulation result shown in Fig. S6(b) is obtained at the center thickness of the loop.

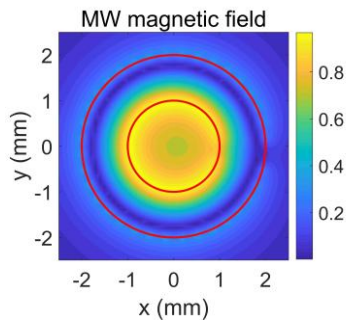

**Fig. S10.** Simulated normalized distribution of the microwave magnetic field. The red annulus (inner radius: 1 mm, outer radius: 2 mm) indicates the position of the loop, with the region between the two red circles representing the current-carrying area. A 0.5 × 0.5 mm$^2$ area at the center marks the location

of the diamond.

## 9. Alignment tolerance analysis of diamond probe

During the adhesive bonding of the diamond, CPC lens, and rod lens in the diamond probe, perfect alignment (complete coaxiality) cannot be guaranteed between the diamond and the CPC lens, nor between the CPC lens and the rod lens. Translational and angular misalignments can occur, causing the optical axes of the three components to deviate from coincidence, which results in a loss of fluorescence collection efficiency. Fig. S11 analyzes the resulting $TCF$ (total collected fluorescence) loss due to such alignment errors. Fig. S11 (a) and (b) illustrate the $TCF$ losses caused by translational misalignment of the diamond relative to the CPC lens, and of the CPC lens relative to the rod lens, respectively. In the analysis of diamond misalignment, the central axes of the CPC lens and the rod lens coincide with the optical axis. In the analysis of CPC lens misalignment, the central axes of the diamond and the CPC lens coincide with each other but are offset from the optical axis, while the central axis of the rod lens remains aligned with the optical axis. Given that the input radius ($r_{in}$) of the CPC lens is 0.51 mm and the half-side length of the diamond base is 0.25 mm, when the lateral displacement of the diamond is less than 0.26 mm (the difference between the CPC input radius and the half-side length of the diamond base), the diamond base remains fully within the CPC input aperture, resulting in a relatively small fluorescence loss ($TCF/TCF_0 > 98.6\%$). However, when the diamond displacement exceeds 0.26 mm, part of the diamond base lies outside the CPC aperture, leading to a significant loss. Similarly, the output radius $r_{out}$ of the CPC lens is 1.5 mm and the radius of the rod lens is 2 mm. When the lateral displacement of the CPC lens exceeds 0.5 mm, a portion of the CPC output surface is no longer coupled to the rod lens, causing a substantial $TCF$ loss. The $TCF/TCF_0$ is over 99.99% when the lateral displacement of the CPC lens is less than 0.5 mm. For alignment tolerance in the range of 50 μm to 100 μm, Fig. S11 shows that the resulting loss in $TCF$ is negligible ($TCF/TCF_0 > 99.99\%$).

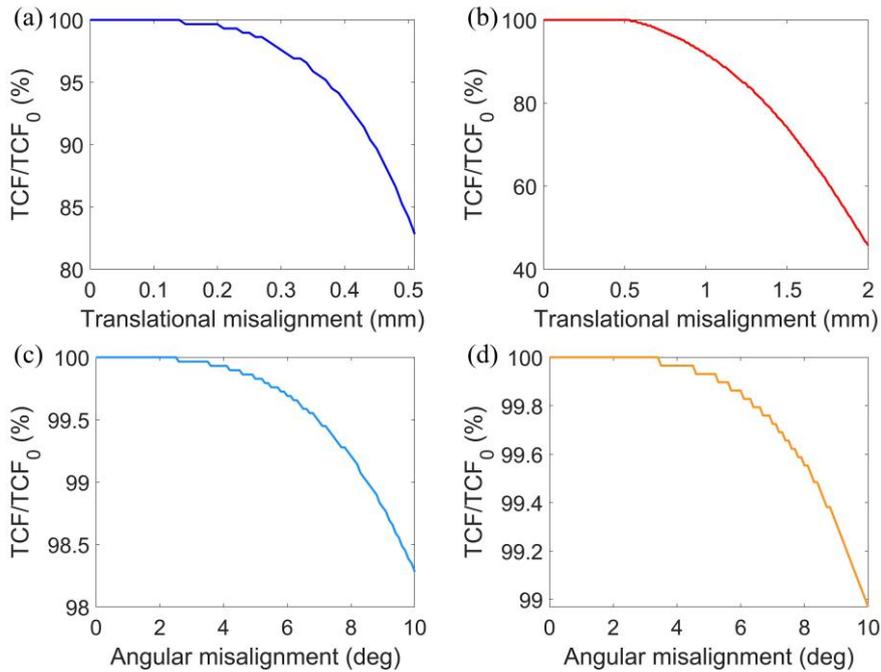

**Fig. S11**. Alignment tolerance analysis of diamond probe. The simulation uses the ideal alignment

$TCF$ ($TCF_0$) as the benchmark. The $TCF$ under all other misalignment conditions is expressed as the ratio relative to $TCF_0$ ($TCF/TCF_0$, relative $TCF$). (a) $TCF$ loss due to translational misalignment of the diamond. (b) $TCF$ loss due to translational misalignment of the CPC lens. (c) $TCF$ loss due to angular misalignment of the diamond. (d) $TCF$ loss due to angular misalignment of the CPC lens.

Fig. S11 (c) and (d) analyze the loss in $TCF$ due to angular misalignment of the diamond and the CPC lens relative to the optical axis, respectively. In the analysis of diamond angular misalignment, the diamond is rotated with respect to the optical axis, while the central axes of the CPC lens and the rod lens remain aligned with the optical axis. In the analysis of CPC lens angular misalignment, the central axis of the diamond coincides with that of the CPC lens, but both are tilted relative to the optical axis, whereas the central axis of the rod lens stays aligned with the optical axis. An angular deviation within 10° represents the controllable range during assembly and is not exceeded in actual bonding. Within this 10° range, the adhesive layer between the optical components helps minimize fluorescence loss into the air. The $TCF/TCF_0$ is over 98.2%.

## 10. Root mean square radius analysis

Fig. S12 illustrates the relationship between the $RMSR$ and the CPC output radius ($r_{out}$), the distance between the probe and the PD ($d$), and the maximum divergence angle in air ($\theta_{am}$). $RMSR$ has a linear correlation with $r_{out} + d\tan\theta_{am}$, with a goodness-of-fit exceeding 99.9%. These results are derived from ray-tracing simulations. Due to the collimating function of the CPC, the divergence angle of the light exiting the probe is always less than a defined limit $\theta_{am}$. Consequently, the farthest light point on the PD does not exceed a distance $r_{out} + d\tan\theta_{am}$ from the center. The $RMSR$ reflects the degree of light concentration. The simulation results show that increasing the CPC size—whether by enlarging its diameter or increasing the divergence angle—leads to a higher $RMSR$ on the PD.

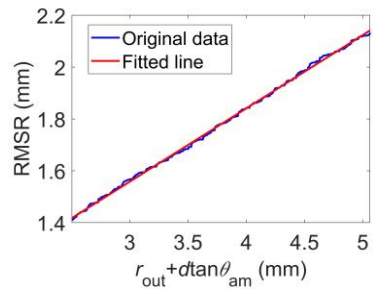

**Fig. S12.** The relationship among the $RMSR$, $r_{out}$, the distance between the probe and the PD ($d$), and $\theta_{am}$. The red line is the fitted linear line of the relationship and the goodness of fit exceeds 99.9%.